\documentclass[12pt,preprint]{aastex}
\begin{document}

\title{The Role of Polycyclic Aromatic Hydrocarbons in Ultraviolet
Extinction. I. Probing small molecular PAHs\altaffilmark{1}}

\author{Geoffrey C. Clayton$^{2}$, Karl D. Gordon$^{3}$, F. Salama$^{4}$, L.J. 
Allamandola$^{4}$, Peter G. Martin$^{5}$, T. P. Snow$^{6}$,
D.C.B. Whittet$^{7}$, A.N. Witt$^{8}$, and Michael J. Wolff$^{9}$}

\altaffiltext{1}{Based on observations made with the NASA/ESA Hubble
Space Telescope, which is operated by the Association of Universities
for Research in Astronomy, Inc., under NASA contract NAS 5-26555.}
\altaffiltext{2}{Department of Physics and Astronomy, Louisiana State 
University, Baton Rouge, LA 70803; Email:
gclayton@fenway.phys.lsu.edu}
\altaffiltext{3}{ Steward Observatory, University of Arizona, Tucson, AZ 
85721 E-mail: kgordon@as.arizona.edu}
\altaffiltext{4}{NASA-Ames Research Center, Space Science Division, MS: 245-6,
Moffett Field, CA 94035-1000 Email: fsalama@mail.arc.nasa.gov,
lallamandola@mail.arc.nasa.gov}
\altaffiltext{5}{Canadian Institute for Theoretical Astrophysics,
University of Toronto, Toronto, Ontario M5S 3H8, Canada Email:
pgmartin@cita.utoronto.ca}
\altaffiltext{6}{Center for Astrophysics and Space Astronomy, 389-UCB,
University of Colorado, Boulder, CO 80309-0389; Email:
tsnow@casa.colorado.edu}
\altaffiltext{7}{Department of Physics \& Astronomy, 
Rensselaer Polytechnic Institute, 
Troy, NY 12180-3590 Email: whittd@rpi.edu}
\altaffiltext{8}{Ritter Observatory, 2801 W. Bancroft, 
University of Toledo, Toledo, OH 43606 Email:
awitt@dusty.astro.utoledo.edu}
\altaffiltext{9}{Space Science Institute, 3100 Marine Street, Ste A353
Boulder, CO 80303-1058 Email: wolff@colorado.edu}

\begin{abstract}
We have obtained new STIS/HST spectra to search for structure in the
ultraviolet interstellar extinction curve, with particular emphasis on
a search for absorption features produced by polycyclic aromatic
hydrocarbons (PAHs).  The presence of these molecules in the
interstellar medium has been postulated to explain the infrared
emission features seen in the 3-13 $\mu$m spectra of numerous sources.
UV spectra are uniquely capable of identifying specific PAH molecules.
We obtained high S/N UV spectra of stars which are significantly more
reddened than those observed in previous studies.  These data put
limits on the role of small (30-50 carbon atoms) PAHs 
in UV extinction and call for further observations to probe
the role of larger PAHs.
PAHs are of
importance because of their ubiquity and high abundance inferred from
the infrared data and also because they may link the molecular and
dust phases of the interstellar medium.  A presence or absence of
ultraviolet absorption bands due to PAHs could be a definitive test of
this hypothesis.  We should be able to detect a 20 \AA\ wide 
feature down to a 3$\sigma$ limit of $\sim$0.02 A$_V$.
No such absorption features are seen other than the well-known 2175 \AA\ bump.

\end{abstract}


\section{Introduction}

Because a large fraction of the photons passing through the interstellar medium (ISM) is
absorbed and re-emitted in the infrared (IR), the intrinsic spectral
energy distribution of reddened astrophysical objects (particularly in
the ultraviolet) cannot be accurately determined without detailed
knowledge of the absorbing medium.  In particular, recent research suggests
that ultraviolet (UV) extinction, particularly the rise toward shorter
wavelengths, may be due to either very small grains or large
molecules, or a combination of both.  It is crucially important for
studies of the chemistry and physics of the ISM to determine which it
is.

While efforts have been made to study interstellar dust in various
Galactic environments at wavelengths from the x-ray to the radio, a
single line of sight is rarely subjected to observations over a wide
wavelength range. This is particularly true for the UV and infrared
 regimes.  In the UV, the amount of extinction due to dust rises
rapidly toward shorter wavelengths and a moderate column density can
easily lead to a paucity of photons.  Consequently, the typical
interstellar sightline used for UV studies has a small column of
dust.  These lines of sight probe the diffuse ISM and avoid all but
the very outer edges of dark or molecular clouds.  IR studies of dust,
on the other hand, probe the dense regions of the ISM and concentrate
on molecular bands of many different solid materials including mixed
molecular ices, silicates and carbon grains associated with dense
cloud environments.  So, to a large extent, studies of interstellar dust
in the UV and IR have been two ``separate worlds.''  Another important
difference between the two is that UV and optical photons probe
transitions between electronic levels and are uniquely capable of
identifying specific atoms and molecules. IR photons, on the other
hand, probe radiative transitions between states associated with molecular 
vibrations
and are characteristic of functional molecular groups rather than
specific molecules. The result of this dichotomy has been models of
dust grains which are strongly biased toward fitting observations in
one wavelength regime or the other.  The absence of lines of sight for
which both UV and IR data are available makes it difficult to
reconcile the constraints derived separately from each wavelength
regime.

Until fairly recently, the popular dust grain models were fairly
simple, having two or three major grain components.  The Mathis model
consisted of power-law size distributions of separate populations of
bare spherical silicate and graphite grains (Mathis, Rumpl, \&
Nordsieck 1977).  The Greenberg model consisted of large grains with
silicate cores surrounded by organic mantles and small carbonaceous
grains, possibly polycyclic aromatic hydrocarbons (PAHs)
\footnote{We use the term PAH generically to describe the
entire class that includes regular PAHs, ionized PAHs, dehydrogenated 
PAHS, PAHs with sidegroups, etc.}
(e.g., Greenberg 1989).  This kind of simple
model has been valuable and gives reasonable fits to the observed
Milky Way dust extinction, polarization and spectral features, but
empirical evidence indicates that interstellar grains are far more
complex.  Attempts are now being made to produce a unified model of
interstellar dust to satisfy observational constraints across all
wavelengths (Mathis 1996, 1998; Li \& Greenberg 1997; Dwek 1997; 
Zubko 1999; Witt 2000).
These much more complex and realistic models of interstellar dust are
now possible through the use of improved observational constraints and
the development of sophisticated (and efficient) numerical techniques.
The abundance of each element available for grain formation provides
important constraints (Sofia \& Meyer 2001).   
Carbon is an important constituent
in all major grain models and may be involved in some or all of the
so-called unidentified infrared emission bands (UIBs) which are
ascribed to PAHs, the diffuse interstellar absorption bands (DIBs),
the near-IR emission continuum, the extended red emission (ERE), the
2175 \AA\ absorption ``bump", and the continuous extinction in the
visible and UV (Mathis 1998; Whittet 2003).

While the models cited above vary quite considerably in the
grain types used, they all associate the production of the UIBs with
PAHs.  Over the past two decades, a combination of observational,
laboratory and theoretical astrophysics has led to the suggestion that
PAH molecules may be the carriers of the UIBs, which consist of a
related series of emission features seen most strongly 
at wavelengths of 3.3, 6.8,
7.7, 8.6, and 11.3 $\mu$m in the spectra of astronomical sources
(Duley \& Williams 1981; 
Puget \& L\'eger 1989; Allamandola, Tielens, \& Barker 1989).
However, critical uncertainties remain regarding the amount of carbon
needed: PAHs are estimated to require from 5\% to almost 20\% of the
available carbon in the ISM (Boulanger et al. 1998; Tielens et
al. 1999).  Some grain models also suggest that PAHs are at least
partially responsible for the 2175
\AA\ bump and the far-UV extinction (Li \& Greenberg 1997).  These claims are 
driven in part by laboratory measurements which show that some PAHs
absorb heavily at far-UV wavelengths, and some even show absorption
bands near the position of the 2175 \AA\ bump (Salama et al. 1996, Salama
1999). However, it is not likely that PAHs alone could produce the
bump.  The strength of the bump is not well correlated with the far-UV
extinction strength, so neither PAHs nor any other single grain-type
can be used to completely explain both extinction components (Mathis
\& Cardelli 1992).

Observations indicate that the carriers of the IR bands must be a
ubiquitous and abundant component of the interstellar medium.  The
bands have been observed in a large variety of sources - the diffuse
ISM, H II regions, reflection nebulae, planetary nebulae, and galactic
nuclei - all characterized by the presence of dust and ultraviolet
radiation (e.g., Tielens 1993; Mattila et al. 1996; Onaka et al. 1996;
Schutte et al. 1998).  PAH emission is most often associated with
photo-dissociation regions (PDRs) which are characterized by fairly
high densities and intense UV radiation fields.  Such densities are
often significantly higher than those seen along most of the diffuse
ISM sightlines studied in the UV.  Two exceptions are the reflection
nebulae, NGC~2023 and NGC~7023.  The central stars of these nebulae
have moderate reddening (E(B-V)=0.4-0.5 mag) and have been studied with
IUE.  Here, PAH emission is seen primarily in filaments in these
nebulae where there are density enhancements.  No unusual structure is
seen in the IUE UV extinction curves of the central stars (HD 37903
and HD 200775; Fitzpatrick \& Massa 1990; Walker et al.
1980).  However, the S/N in IUE data is not very high and, as we
describe below, the predicted UV PAH band strengths are small for
these values of E(B-V).  
Other attempts to detect and identify absorption features in the UV are 
detailed in Tripp, Cardelli, \& Savage (1994).
Successful detection of the UV electronic
transitions would allow for the identification of specific species,
something that is not possible through observations of the IR
vibrational transitions, which probe only the presence of particular
molecular bonding arrangements (e.g., aromatic vs.  aliphatic
hydrocarbons).

\section{Observations and Analysis}

We selected reddened stars whose dust columns are significantly higher
than those of previous UV studies.  Specifically, we observed three
reddened stars, listed in Table 1, having E(B-V) between 1.3 and 1.7 mag.
All three of these sightlines pass through relatively dense clouds;
two, HD 194279 and HD 229059, in the Cyg OB2 cloud and one, HD 283809,
in the Taurus dark cloud.  They all show evidence for the 3.4
\micron\ feature in absorption (Smith, Sellgren, \& Brooke 1993;
Pendleton et al. 1994; Gordon et al. 2003).  
The 3.4 \micron~feature is associated with a
C-H stretching vibration in aliphatic hydrocarbons that are believed
to be part of organic refractory grains made of aromatic and aliphatic
units (Pendleton \& Allamandola 2002).  Relatively unreddened stars
with spectral types well-matched to the reddened stars were also
observed.  The unreddened standards were selected to match the
reddened stars to within one half of a spectral subtype (B1.5 versus
B1).  This minimizes the noise due to mismatch between the stellar
features of the reddened and unreddened stars. There was some difficulty in 
finding lightly reddened stars of matching spectral types that were faint
enough to observe with STIS without saturating the detector. For instance,
we were
forced to choose HD 151805 even though it has a significant reddening.

The five stars in this program were all observed with STIS/HST and the
MAMA detectors with both the G140L (1150-1730 \AA) and G230L
(1570-3180 \AA) gratings. These low resolution gratings provide
two-pixel resolutions of 1.2 \AA\ (G140L) and 3.2 \AA\ (G230L).  The
observation dates, exposure times and apertures used are listed in
Table~\ref{tab_data} for each of the five stars. HD 283809 and HD
229059 were observed through the 52x2 slit and the data were reduced
using the On-The-Fly calibration.  However, the count rates for the
other three stars required that a neutral-density slit be used.
Neutral-density-slit modes are unsupported but good calibrations were
possible using observations of the white dwarf, GD 153, made through
both the standard slit (52x2) and the neutral-density slits
(31x0.05NDB, 31x0.05NDC), which were available in the MAST archive.
The GD 153 observations are also listed in Table~\ref{tab_data}.  The
conversions between the neutral-density slits and the standard slit
were derived by dividing the 52x2 aperture spectrum by 31x0.05NDB and
31x0.05NDC spectra of GD 153.  The ratio spectra were smoothed with a
31-pixel-width boxcar.  These ratio spectra were essentially flat with
smooth variations across the wavelength range of around 25\%.  These
smoothed spectra were used to convert the measured counts for the
neutral-density aperture observations to calibrated spectra.

The calibrated G140L and G230L spectra were coadded using weights
defined by the pipeline-produced uncertainties to produce a single
spectrum for each star.  Extinction curves were constructed using the
standard pair method (e.g., Massa, Savage, \& Fitzpatrick 1983).
The estimated uncertainties in the extinction curve contain terms that depend
on the broadband photometric uncertainties as well as uncertainties in
the STIS fluxes.  For details of our error analysis, the reader is
referred to Gordon \& Clayton (1998).  The uncertainty term which most
affects our ability to search for new structure in extinction curves
is that due to random uncertainties in the flux measurements.
Uncertainty terms which affect the overall level of the extinction
curve (e.g., B and V magnitude uncertainties) do not mask the presence
of structure.

The extinction curves for the reddened/comparison star pairs, HD
229059/HD 151805 and HD 283809/HD 51038, are shown in the upper panels
of Figures~\ref{fig_hd229} and \ref{fig_hd283}, respectively.  We have
not shown the extinction curve for the HD 194279/HD 151805 pair as it
suffers from significant spectral mismatch (mainly luminosity
mismatch).  This results in an extinction curve dominated by
mismatches in stellar lines which is unsuitable for our purposes.  
Therefore, the HD 194279 sightline is not included in the discussion below.
The
STIS/HST spectra of the stars are plotted in the lower panels of
Figures~\ref{fig_hd229} and \ref{fig_hd283} for HD 229059/HD 151805 and
HD 283809/HD 51038, respectively.

In order to look for new structure in the extinction curves, we made
fits to the full UV through near-IR extinction curves.
These fits included the known structures of the 2175 \AA\ bump and
far-UV rise. The most commonly used fit to UV extinction curves is the
Fitzpatrick \& Massa (1990) parameterization.  
However, the Fitzpatrick \& Massa fitting function is not
valid to the red of 2700 \AA\ (Fitzpatrick 2002, personal
communication).  Therefore, we adopted a function similar to that of
eq.~20 of Pei (1992).  This fitting function covers the entire range
from X-ray to far-IR wavelengths and consists of 6 terms with a
total of 19 free parameters.  Restricting the function to just the
near-IR to UV wavelength range reduces the fit to 3
terms (Background, Far-UV, and 2175 \AA\ terms) with 9 free parameters.  Pei
(1992) used this function to fit, by eye, the average extinction
curves in the Milky Way and Magellanic Clouds and found it to be
accurate to $0.1A_V$.  We used a numerical curve fitting routine in
IDL (mpfitfun) to determine the 9 free parameters and find that the
fitting function can reproduce the broad structure of the
extinction curve quite accurately.

The fits were subtracted from the measured extinction curves and the
resulting residuals are plotted in the middle panel of
Figures~\ref{fig_hd229} and \ref{fig_hd283}.  The uncertainties which
limit our ability to detect new extinction curve structure are those
associated with random uncertainties in the ultraviolet flux
measurements and, to a lesser extent, spectral mismatch uncertainties.
The random uncertainties are plotted in the middle panel of
Figures~\ref{fig_hd229} and \ref{fig_hd283}.  The uncertainties
associated with spectral mismatches will occur at specific wavelengths
(coincident with known stellar lines) and must be evaluated on a
case-by-case basis.

The residuals were examined for previously unseen features above the
uncertainties.  A positive feature in the residuals would signify
excess absorption.  In the HD~229059 extinction curve, there are
significant residuals at approximately 4.6, 6.5, and 7.2
$\micron^{-1}$.  The 4.6 $\micron^{-1}$ residual is at the center of
the 2175~\AA\ bump and is due to the absence of stellar flux in the
middle of the 2175 \AA\ feature where the absorption is strongest (see
Figure~\ref{fig_hd229} bottom panel).  The resulting large
uncertainties in the extinction curve for HD~229059 cause the bump fit
parameters to be less certain.  The oscillations seen in the residuals
from 4 to 5 $\micron^{-1}$ for this extinction curve are caused by the
uncertain bump fit parameters.  The 6.5 and 7.2 $\micron^{-1}$
residuals are due to known stellar lines, C~IV and Si~IV, which are
difficult to match exactly for supergiants (Cardelli, Sembach, \&
Mathis 1992).  In the HD~283809 extinction curve, there is a residual
feature at approximately 6.5 $\micron^{-1}$.  This apparent feature is
caused by a luminosity mismatch (Massa et al. 1983) .  
Well-known interstellar absorption lines such as Mg II $\lambda$$\lambda$2798, 2803,  Fe II $\lambda$$\lambda$2344, 2382, 2600, C II $\lambda$1334, 
O I $\lambda$1302 and Si II $\lambda$$\lambda$1260, 1304 are clearly present 
in the spectra of the reddened stars. These lines cause 
narrow mismatch features between the reddened and unreddened stars.
Thus, we do not find any new features in the extinction curves
of HD~229059 or HD~283809 at a 3$\sigma$~limit 
of $0.02A_V$.
See further quantification below.

\section{Discussion}

The intensity ratios in the UIB emission imply that PAH molecules with
20-200 C atoms are responsible for most of the emission from the ISM (Allamandola et al. 1989; Schutte,
Tielens \& Allamandola  1993, 
Salama et al. 1996; Allain, Leach, \& Sedlmayr 1996a,b).  
More recent modeling work, which had the benefit of having a much more 
extensive ISO data base available for comparison, also suggests minimum numbers
of carbon atoms in PAH molecules of order 20 - 30 (Verstraete et al.  2001; Pech, Joblin \& Boissel 2002; Peeters et al. 2002).
Indeed, 
comparing the astronomical spectra with the laboratory spectra of 
mixtures of larger PAHs seems to do somewhat better.
The detection of the IR
bands in reflection nebulae, where the starlight energy densities are
quite insufficient to produce equilibrium thermal emission, 
has led to the point of view that they are excited in a
fluorescence process (Allamandola \& Norman 1978; Allamandola, Greenberg, 
\& Norman 1979; Sellgren 1984, Li \& Draine 2002).  Under the
PAH hypothesis, the fluorescence is induced by the absorption of UV
and visible photons, with the IR emission features being produced by
vibrational relaxation of the excited molecule.
PAH molecules and ions have absorption bands
that span the UV-visible and near-IR ranges (e.g., Chillier et al. 1999).

The photodestruction of PAHs under interstellar conditions
has been modeled (Jochims et al. 1994; Allain et al. 1996a,b;
LePage, Snow, \& Bierbaum 2003).
The first group find a critical size in the range of 30 - 40 carbon atoms,
below which PAHs are likely being destroyed. The second group concludes
that the critical size is more likely around 50 atoms. LePage et al. (2003) suggest that PAHs with fewer than 15-20 atoms are mostly destroyed, those
with 20-30 atoms are dehydrogenated, and larger PAHs are largely ionized.
The apparent scarcity of small PAHs in the
ISM is supported by the absence of any detection of dust luminescence
associated with {\it neutral} PAHs
shortward of 5400 \AA\ in the spectra of reflection nebulae, which are known
sources of UIB emission (Donn et al. 1989; Rush \& Witt 1975).
Other studies have suggested dominance by both neutral and ionic 
species in the ISM (Salama et al.
1996; Weingartner \& Draine 2001).

UV absorption spectra of
isolated PAH species at the low temperatures characteristic of the ISM
have been obtained in the laboratory for PAHs ranging from naphthalene
($C_{10}H_{8}$) up to dicoronene ($C_{48}H_{20}$) (e.g., Salama,
Joblin \& Allamandola 1995; Salama et al. 1996; Chillier et al. 1999;
Salama 1999; Ruiterkamp et al. 2002). These data constitute a
representative dataset for the UV and visible spectra of PAHs.
Figure~\ref{fig_pah} shows selected examples.

\begin{figure}
\epsscale{0.65}
\plotone{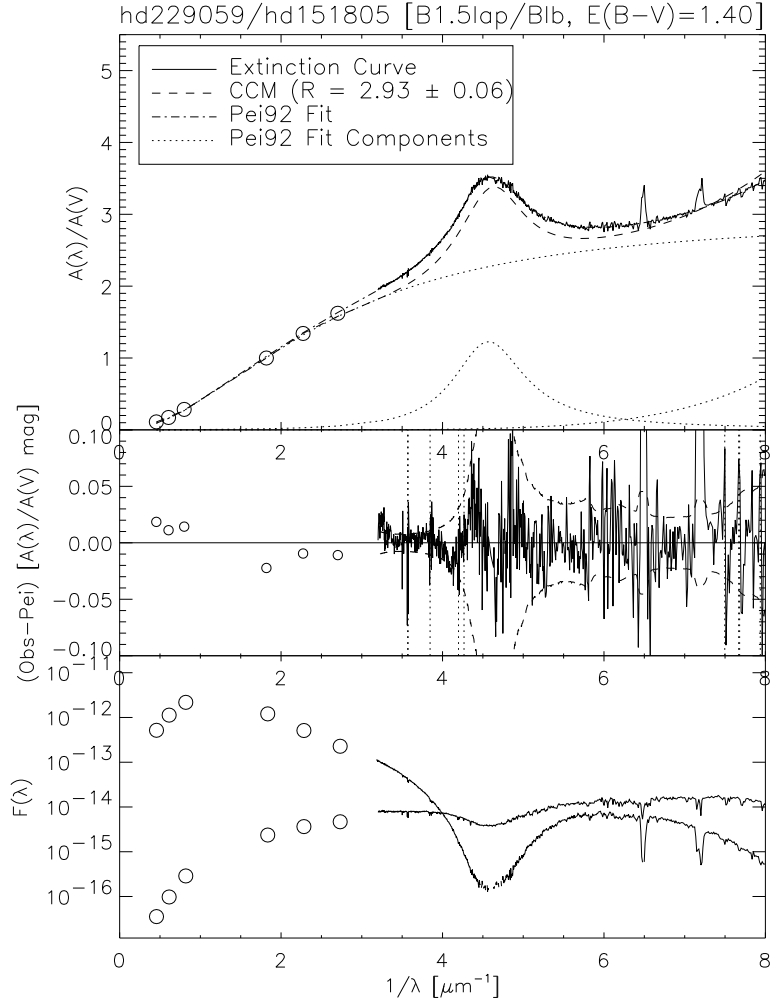}
\caption{Top panel: The measured extinction curve for the HD 229059/HD
151805 pair is shown along with the Pei (1992) fit and the Cardelli,
Clayton \& Mathis (1989) curve for the adopted $R_V$ value.  The $R_V$
value is 3.60 and was measured JHK photometry from Leitherer \& Wolff
(1984) for HD~229059 and 2MASS (Cutri et al. 2000) for HD~151805.
Middle panel: The deviations from the Pei fit are shown along with the
$3\sigma$ detection limits (dashed line) due to random flux
measurement uncertainties.  The data are plotted every pixel; the
two-pixel resolution corresponds to 1.2 \AA\ in the far-UV and 3.2
\AA\ in the near-UV, respectively.  The vertical dotted lines mark the
locations of interstellar absorption lines which cause narrow mismatch
features between the reddened and unreddened stars.  Bottom panel: The
reddened (HD~229059) and comparison (HD~151805) star spectra are
plotted. The comparison star spectrum has been scaled by a constant to
fit on the plot with the reddened star spectrum. \label{fig_hd229}}
\end{figure}

\begin{figure}
\epsscale{0.65}
\plotone{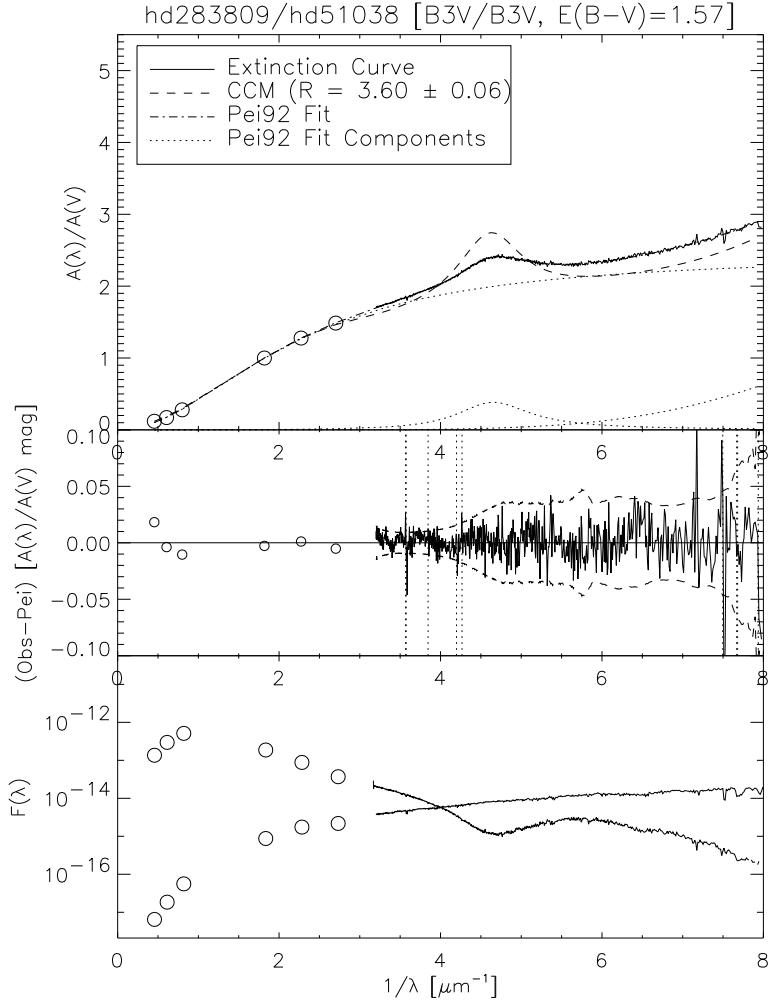}
\caption{Top panel: The measured extinction curve for the HD 283809/HD
51038 pair is shown along with the Pei (1992) fit and the Cardelli,
Clayton \& Mathis (1989) curve for the measured $R_V$ value. The $R_V$
value is 3.60 and was measured from 2MASS photometry (Cutri et al. 2000)
for both
stars. Middle panel: The deviations from the Pei fit are shown along
with the $3\sigma$ detection limits (dashed line) due to random flux 
measurement uncertainties.  
The data are plotted every pixel; the two-pixel resolution corresponds to 
1.2 \AA\ in the far-UV and 3.2 \AA\ in the near-UV, respectively.
The vertical dotted lines mark the locations of interstellar absorption lines
which cause narrow mismatch features between the reddened and unreddened stars.
Bottom panel: The reddened (HD~283809) and 
comparison (HD~51038) star spectra are plotted. The comparison star 
spectrum has been scaled by a constant to fit on the plot with the reddened 
star spectrum. \label{fig_hd283}}
\end{figure}

\begin{figure}
\epsscale{0.8}
\plotone{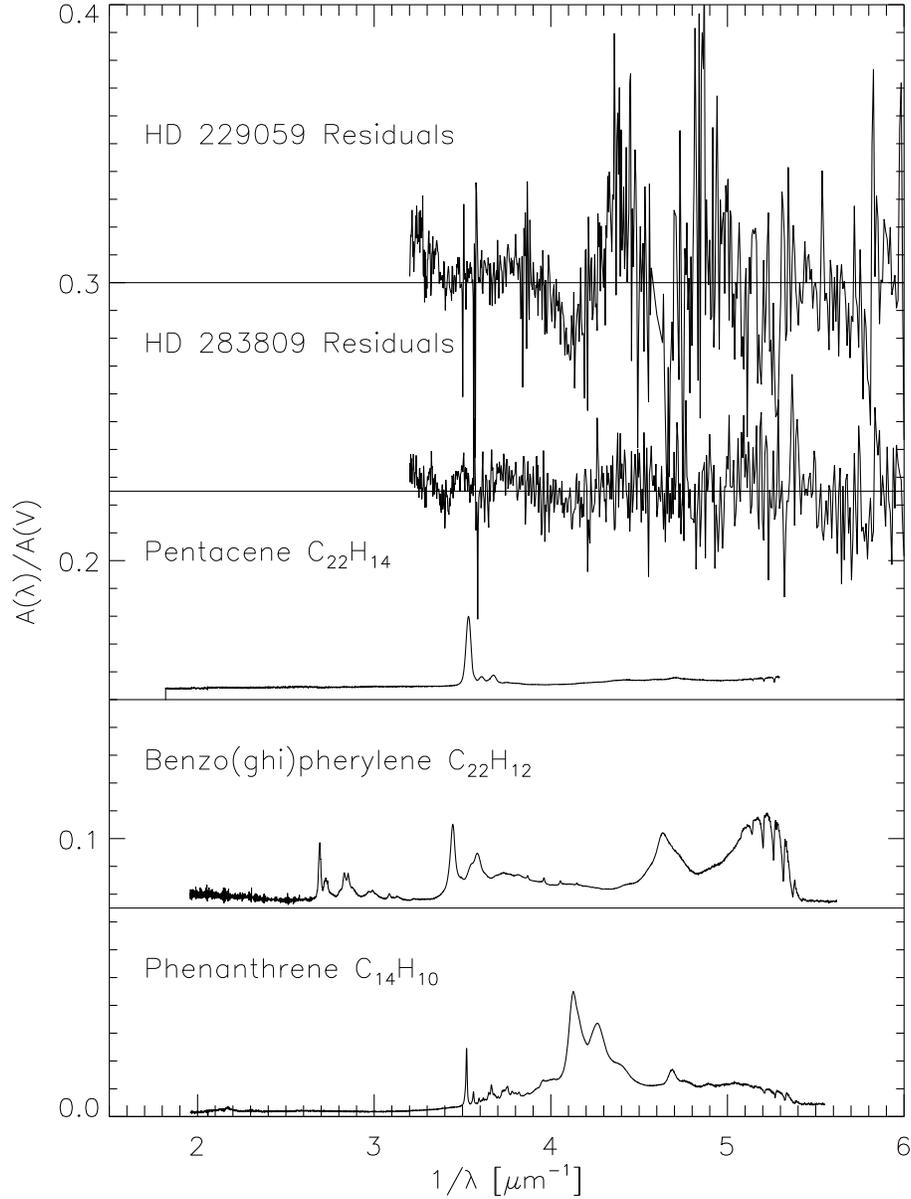}
\caption{Ultraviolet absorption spectra of three representative PAHs are
plotted along with the Pei (1992) fit residuals for HD 229059 and
283809.  The laboratory spectra are of PAHs isolated in inert-gas
matrices at 5~K.  The strongest PAH absorption feature are scaled to
0.045, 0.03, and 0.03~A$_V$ for Phenanthrene, Benzo(ghi)perylene, and
Pentacene, respectively.   The spectra have been shifted 
vertically relative to one another by units of 0.075 to 
allow for easy comparison.
\label{fig_pah} }
\end{figure}

Earlier attempts to detect such features with IUE and Copernicus
failed because the S/N was too low in the stellar spectra and the dust
column was too small (Snow, York, \& Resnick 1977; Seab \& Snow 1985).
The accuracy of the residuals of the two extinction curves plotted
in A($\lambda$)/A$_V$ space in Figures \ref{fig_hd229} and
\ref{fig_hd283}, is inversely proportional 
to the amount of extinction and also depends on the quality of the
spectral-type match.  Of the two residual curves, HD 283809 has the
best combination of high reddening and good spectral match.  
The
residual curve for HD~229059 is not as smooth, but is good enough to
put a similar limit on broad features.  
For narrow (a few \AA, near the resolution limit of our data) stellar photopheric-type lines, the 3$\sigma$ 
limit is $\sim$0.03-0.04 A$_V$. 
Laboratory spectra show that PAH features are considerably broader than our
resolution. 
The detectability of a feature rises as it becomes broader, 
the detection limit goes down as the square root of the width of the 
feature. For example, we should be able to detect a 20 \AA\ wide 
feature\footnote{ Recent gas-phase data indicate that PAH bands 
are broadened by a factor
of 4 -5 in solid matrices (Biennier et al. 2003).}
 down to a 3$\sigma$ limit of $\sim$0.02 A$_V$.
Looking at Figures 1 and 2, clearly,
there are no 
broad PAH-like features with depths greater than $\sim$0.02 A$_V$.

From this we can estimate the upper limit on the abundance of a
PAH molecule. The absorption cross-section
of the naphthalene molecule averaged over the strong 2120 \AA\ band is
about 4 x 10$^{-16}$ cm$^2$.  The strong bands of the other PAH
species have comparable strengths.  We assume the average Galactic
gas-to-dust ratio, N$_H$/E(B-V) = 5.8 x 10$^{21}$
atoms~cm$^{-2}$~mag$^{-1}$ (Bohlin, Savage, \& Drake 1978).  So with
A$_V$ = N$_H
R_V$/5.8 x 10$^{21}$ mag, and taking 2.5 x 10$^{-16}$ cm$^2$ as a
typical band strength, we find that the average optical depth in a
strong PAH band would be: $\tau$ = 2.5 x 10$^{-16}$(N(PAH)/N(H))(A$_V$/R$_V$)5.8 x 10$^{21}$.  
The upper limit on our detection is $\tau <$ (1.086)0.02A$_V$ so 
N(PAH)/N$_H$ $<$ (1.086) 0.02R$_V$ /((2.5 x 10$^{-16}$)(5.8 x 10$^{21}$)) = 
4.6 x 
10$^{-8}$(R$_V$/3.1).
We assume here for simplicity that N$_H$/E(B-V) is a constant for all R$_V$ (see Kim \& Martin 1996).

The abundance of a single species of PAH could be given by
N(single species of PAH)/N$_H$ =(the Cosmic carbon abundance, C/H) x 
(the fraction of C in all PAHs) x (the fraction of C in all PAHs in a single 
PAH species)/(the number of C atoms in this PAH molecule). The cosmic 
abundance of carbon is not well known (e.g., Sofia \& Meyer 2001; Prieto, 
Lambert \& Asplund 2002). We will assume C/H = 3 x 10$^{-4}$. 
As discussed above, it has been estimated that 5 - 20\%\ of the carbon is in 
the form of
PAHs, and that PAHs with 20 - 200 C atoms best fit the UIB
emission. So, using these ranges, assuming that the fraction of this one PAH
species is 0.10, that the fraction of available C in PAHs is 0.05,
and the number of C atoms in 
the PAH species is 20 then N(single species of PAH)/N$_H$ = 7.5 x 10$^{-8}$. 
For a larger PAH species of 200 C atoms, we get N(single species of PAH)/N$_H$
 = 7.5 x 10$^{-9}$. For a fraction of available C in all PAHs of 0.20, we get 
N(single species of PAH)/N$_H$ = 
3.0 x 10$^{-7}$ for a 20-atom PAH species and 3.0 x 10$^{-8}$ for a 200-atom 
species, respectively. 
>From these estimates, we would expect that 20-atom PAH species should be
above our detection limit, while 200-atom species would only be detectable 
if they are very abundant. 

The predicted UV
spectral signatures of PAHs while weaker than the well-known 2175
\AA\ absorption feature, nevertheless, were expected to be much
stronger in the highly reddened sightlines of our sample than in those
previously studied.  However, individual  PAH molecules whose
absorption cross-sections are typical of those measured in the
laboratory do not seem to be present in numbers large enough to
produce measurable UV absorption bands.  
As described above, PAHs with fewer than 30-50 carbon atoms, thought to have
strong UV absorptions, may be destroyed by the 
ambient interstellar radiation field.
Another possibility is that
many PAH species are present in small numbers each producing a small amount
of absorption throughout the UV wavelength range, i.e., 
the absence of detectable PAH absorption does not, in and of itself,
prove an absence of PAH molecules in the ISM.
There is nothing known about the
sightlines toward the Taurus cloud (HD 283809) or Cyg OB2 (HD 229059)
that would imply that the abundance of PAH molecules 
would be low along these sightlines.  
The UIB emission
seems to be ubiquitous in the Galaxy, even being present in the
diffuse ISM (e.g., Chan et al. 2001).

These new STIS data provide a critical test of the PAH model.  The
results of this study serve to put upper limits on the abundance of
individual PAH species in the ISM.  They provide, in particular,
constraints on the abundance of smaller  PAHs (i.e., containing
fewer than 30-50 carbon atoms) in these lines of sight.  They also
call for higher S/N observations (at the $0.01A_V$ level or better) in
the UV to search for absorption associated with larger molecular PAHs
that are now thought to be representative of the interstellar PAH
population.  

This work was supported by STScI grant GO-08670.01-A.

\clearpage

\begin{deluxetable}{lllllll}
\tablewidth{0pt}
\tablecaption{STIS Program Stars \label{tab_data}}
\tablehead{ \multicolumn{1}{c}{Star} & \multicolumn{1}{c}{Sp.T.} & 
  \multicolumn{1}{c}{E(B-V)} & \multicolumn{1}{c}{Obs.\ date (UT)} &
  \multicolumn{1}{c}{Grating} & \multicolumn{1}{c}{Exposure (s)} &
  \multicolumn{1}{c}{Aperture} }
\startdata
HD 194279 & B1.5Ia  & 1.22    & 2001 April 22   & G140L & 850  & 31x0.05NDB \\
          &         &         & 2001 April 22   & G230L & 757  & 31x0.05NDB \\
HD 229059 & B1.5Ia  & 1.73    & 2001 April 18   & G140L & 5318 & 52x2 \\
          &         &         & 2001 April 21   & G230L & 5311 & 52x2 \\
HD 283809 & B3V     & 1.62    & 2000 December 8 & G140L & 5194 & 52x2 \\
          &         &         & 2000 December 9 & G230L & 5187 & 52x2 \\ 
\tableline
HD 51038  & B3V     & 0.05    & 2001 February 2 & G140L & 800  & 31x0.05NDC \\
          &         &         & 2001 February 2 & G230L & 714  & 31x0.05NDC \\
HD 151805 & B1Ia    & 0.33    & 2001 April 3    & G140L & 850  & 31x0.05NDB \\
          &         &         & 2001 April 3    & G230L & 757  & 31x0.05NDB \\ 
\tableline
GD 153\tablenotemark{a}    & WD      & \nodata & 1997 July 13    & G140L & 187  & 52x2 \\
          &         &         & 1997 July 13    & G230L & 187  & 52x2 \\
          &         &         & 1997 July 13    & G140L & 187  & 31x0.05NDB \\
          &         &         & 1997 July 13    & G230L & 279  & 31x0.05NDB \\
          &         &         & 1997 July 13    & G140L & 437  & 31x0.05NDC \\
          &         &         & 1997 July 13    & G230L & 714  & 31x0.05NDC \\
\enddata
\tablenotetext{a}{The observations of GD 153 were used to calibrate the STIS 
neutral
density filters.}

\end {deluxetable}

\clearpage

\end{document}